\documentclass[journal=jpccck,manuscript=article,layout=twocolumn]{achemso} %
\usepackage[version=3]{mhchem} 
\usepackage{graphicx}
\usepackage{dcolumn}
\usepackage{bm}

\usepackage{mathtools}
\usepackage{amssymb}
\usepackage{amsmath}
\usepackage{amsthm}
\usepackage{graphicx}
\usepackage{bm}
\usepackage[dvipsnames]{xcolor}
\usepackage{epsfig}
\usepackage{amsfonts}
\usepackage{bbold}
\usepackage{multirow}
\usepackage{footnote}
\usepackage{blkarray}
\makesavenoteenv{tabular}
\makesavenoteenv{table}
\usepackage{multicol, blindtext}
\usepackage{tablefootnote}
\usepackage[]{appendix}



\author{C. Wolf}
\affiliation{Center for Quantum Nanoscience, EWHA Womans University, Seoul, Republic of Korea}
\altaffiliation{Contributed equally to this work}
\author{F. Delgado}
\affiliation{Departamento de F\'isica, Universidad de La Laguna 38203,  Tenerife, Spain}
\altaffiliation{Contributed equally to this work}
\email{fernando.delgado@ull.edu.es}
\author{J. Reina}
\author{N. Lorente}
\affiliation{Centro de F\'isica de Materiales, Centro Mixto CSIC-UPV/EHU, Paseo Manuel de Lardizabal 5, E-20018 Donostia-San Sebasti\'an, Spain }

\title{Efficient Ab-initio Multiplet Calculations for Magnetic Adatoms on MgO}

\abbreviations{ESR, STM, DFT}

\begin{document}

\begin{tocentry}
\includegraphics[width=80mm]{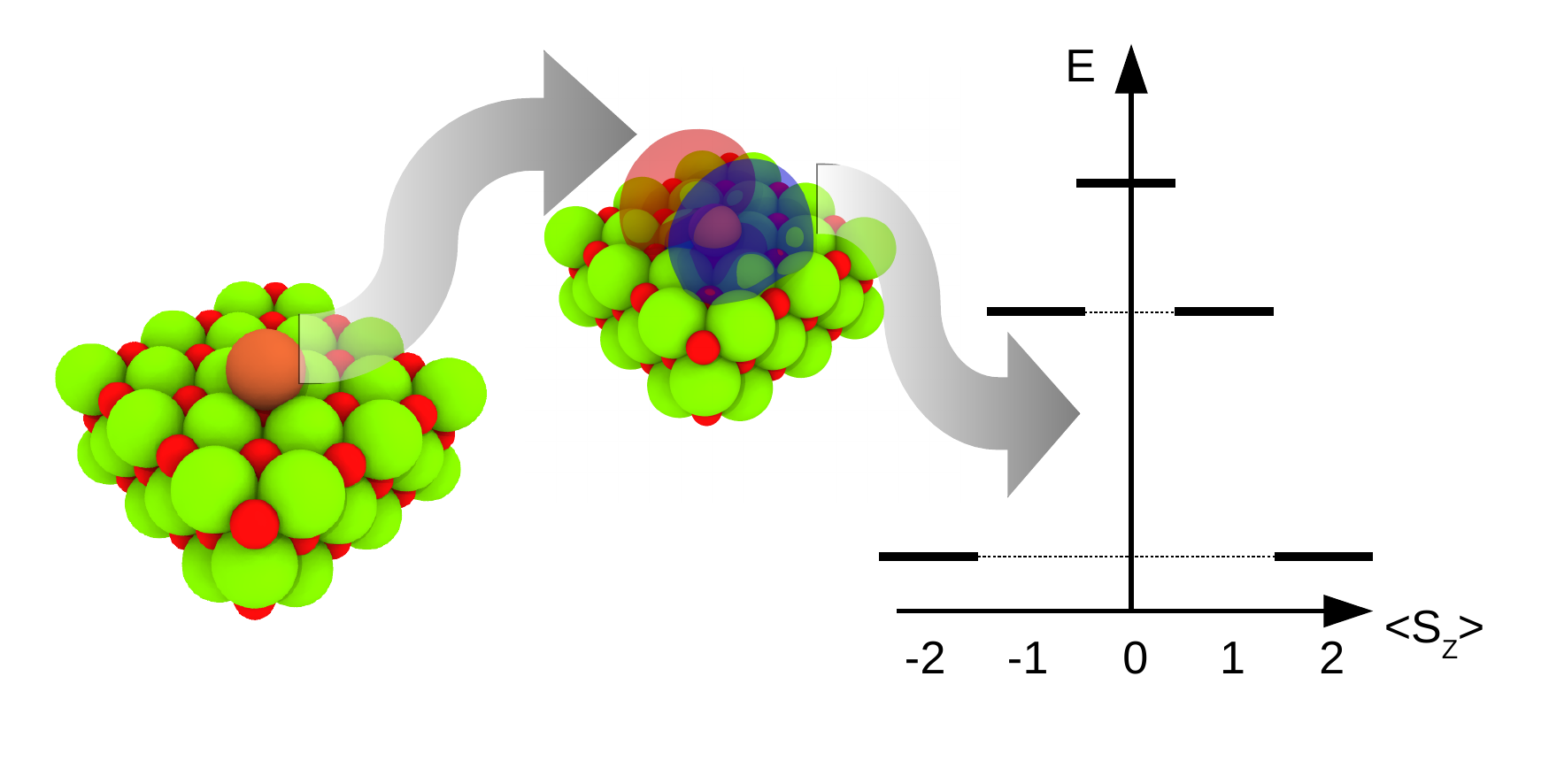}
\end{tocentry}

\begin{abstract}
Scanning probe microscopy and spectroscopy, and more recently in combination with electron spin resonance, have allowed the direct observation of electron dynamics on the single-atom limit. The interpretation of data is strongly depending on model Hamiltonians. However, fitting effective spin Hamiltonians to experimental data lacks the ability to explore a vast number of potential systems of interest.
By using plane-wave density functional theory (DFT) as starting point, we 
build a multiplet Hamiltonian making use of maximally-localized Wannier functions. The Hamiltonian
contains  spin-orbit and electron-electron interactions needed to obtain the relevant
spin dynamics. The resulting reduced Hamiltonian is solved
by  exact diagonalization. We compare three prototypical cases of 3d transition metals Mn (total spin $S=5/2$), Fe ($S=2$) and Co ($S=3/2$) on MgO with experimental data and find that our calculations can accurately predict the spin orientation and anisotropy of the magnetic adatom. Our method does not rely on experimental input and permits us to explore and predict the fundamental magnetic properties of adatoms on surfaces.
\end{abstract}

\maketitle

\section{Introduction}
The quantum spin states of magnetic impurities on surfaces are candidates for magnetic data storage and quantum bits (qubits).\cite{Natterer2017, Baumann_Donati_prl_2015, Donati_Rusponi_science_2016} The properties of such a magnetic state are largely determined by the magnetic anisotropy energy (MAE), which is typically of the order of a 1 meV/atom for transition metal (TM) bulk structures and thin films, but can reach up to several tens of meV/atom in the case of isolated TM-adatoms on thin decoupling layers.\cite{Donati_Dubout_prl_2013,Baumann_Donati_prl_2015,
Rau2014a,Gallardo_2019,rejali2019independent} 

Predicting the MAE of magnetic adatoms on surfaces \textit{ab-initio}  is a difficult task as calculations based on density functional theory (DFT) systematically underestimate the magnitude and sometimes even fails to predict the preferred orientation of the magnetic moment.\cite{Eschrig2005, Bonski2010} This has been addressed by applying a (often empirical) Hubbard U correction,\cite{Ou2015, Mazurenko_Iskakov_prb_2013} or adding a term accounting for orbital polarization.\cite{Eriksson1990} 
A frequently used option is to fit an effective spin Hamiltonian to experimental data and extract anisotropy parameters. This situation is unsatisfying and a method is required that allows one to systematically search for systems of large MAE (and thus potentially long $T_1$ lifetimes) without relying on experimental observations.

An interesting alternative approach is multiplet ligand-field theory (MLFT) using Wannier orbitals,\cite{Haverkort2012a} where the many body nature of the problem is described in a local basis set, which spans the low energy band structure of the system. This method is entirely compatible with the ab-initio concept. There are several related proposals on how to construct the local basis. For example, Haverkort {\em et al.}\cite{Haverkort2012a} used a linear combination of Wannier orbitals centred on the TM and the neighbouring atoms to described the crystal and ligand field, whereas Ferr\'on {\em et al.} employed a basis of maximally localized Wannier functions (MLWFs).\cite{Ferron2015,Ferron_Lado_prb_2015}  
While the former may be more efficient in reducing the Hilbert space of the MLFT calculations, especially when the environment is not very ionic, the use of MLWFs better complies with the assumptions of atomic orbitals made in the derivation of the multiplet.\\

Here we present a method that takes advantage of the strong points of both methods. On one hand, we use MLWFs to describe the one-electron structure of the TM orbitals in their environment. On the other hand, we use a linear combination of MLWFs that efficiently reproduce the low-energy density functional band structure of the supporting substrate in the energy window of interest. Our method is able to accurately calculate the preferred orientation of the magnetization axis, the spin-excitation energy (comparable to inelastic electron tunneling spectroscopy, IETS),\cite{Heinrich2004} the magnitude of the MAE and spin and orbital contribution to the latter conserving the predictive character of ab-initio methods. Further the proposed method allows us to study the individual contribution of crystal and ligand field (CF), spin-orbit coupling (SOC) and Zeeman splittings by an external magnetic field. We compare our calculations with the two experimentally well-studied TM ions, Fe and Co, and make predictions for the case of Mn, where spin-transitions on MgO have so far not been observed. \\

\section{Method}
The goal is to describe a magnetic surface system as typically explored in scanning tunneling microscopy (STM) experiments where a magnetic impurity is absorbed on an insulating surface. The method should (i) not rely on any experimental input and (ii) accurately describe the electron correlation on the localized states of interest (here the 3d manifold of a TM). To capture the physics of a magnetic system we propose the following Hamiltonian:
\begin{equation}
    H= \lambda_{\rm Cou}H_{{\rm Coul}}+\lambda_{\rm CF}H_{\rm CF}+\lambda_{\rm SO}H_{\rm SO}+H_{\rm Zeem}.
\label{eq:htot}
\end{equation}
The terms in  eq \ref{eq:htot} are: electron-electron interaction (Coul), crystal and ligand fields (CF), spin-orbit coupling (SO) and Zeeman splitting due to an external field (Zeem). The dimensionless $\lambda$-parameters allow us to control the relative strength of each contribution by varying them from [0,1], where $\lambda_x=1$ corresponds to the real physical system. 

The above Hamiltonian is solved by exact diagonalization in a reduced many-body Hilbert subspace. An important feature of the present method is that calculations can be strongly simplified if the TM orbitals are \textit{atomic-like}, especially for the Coulomb and SOC terms. The orthogonality of the MLWFs together with their similarity to atomic-orbitals, associated with their small spread, makes them a good choice as a basis set.
\\

\subsection{DFT calculations and local Wannier basis}
The starting point of our method is a DFT calculation using a plane-wave basis set and pseudopotentials. The Bloch states $|\psi_{n\textbf{k}}\rangle$ are generalized Fourier-transformed into real space Wannier functions (WFs) by 
\begin{equation}
    |\textbf{R}n\rangle=\frac{V}{(2\pi)^3}\int_{BZ}d\textbf{k}e^{-i\textbf{k}\cdot\textbf{R}} \sum_m U^{(\textbf{k})}_{mn}|\psi_{m\textbf{k}}\rangle,
    \label{eq:wannier}
\end{equation}
where \textbf{R} is a real-space lattice vector, \textit{BZ} denotes the first Brillioun zone, \textit{V} is the real-space volume of the unit cell and \textit{n}, \textbf{k} are band index and wave vector, respectively. The unitary transformation $U^{(\textbf{k})}_{mn}$ is used to maximize the localization of the Wannier function.\cite{Marzari1997} This calculation is carried out in Wannier90 which implements MLWFs.\cite{Marzari2012, Pizzi2019}\\
The resulting Wannier basis qualitatively satisfies the aforementioned conditions and has dimensions of the order $N_W=$100. Since eq~(\ref{eq:htot}) will be solved by exact diagonalization, a reduction of the single-particle basis is required. We achieve this by using linear combination of Wannier orbitals (LCWO) of the substrate, as described in detail below. This step leads to a significant improvement over the Wannier function basis used previously~\cite{Ferron_Lado_prb_2015}. In addition, it preserves the atomic-like character of the TM orbitals (for a visual representation of the Wannier orbitals, see Figure~\ref{fig:mlwfs}). 

\subsection{Reduced Wannier Hamiltonian \label{reduce}}
 After ensuring that our supercell employed in DFT is large enough we retain only the ${\bf R=0}$ part of the Wannier Hamiltonian and neglect the hybridization with neighboring supercells. The resulting-single particle Hamiltonian has dimension $N_W\times N_W$ and can be arranged in the form
\begin{equation}
{\bf H}_W=
\left[
\begin{array}{cc}
 {\bf H}_{sd} &  {\bf V}_{sd,\alpha}\\
 {\bf V}_{\alpha,sd} &  {\bf H}_{\alpha}\\
\end{array}
\right],
\label{eq:hmat}
\end{equation}
where ${\bf H}_{sd}$ (${\bf H}_{\alpha}$) corresponds to the Hamiltonian of the TM (surface states) and ${\bf V}_{sd,\alpha}$ is a hopping matrix. ${\bf H}_{sd}$ is spanned by the bases of $s$ and $d$-like MLWFs, and has dimension $N_{sd}=6$. The dimension $N_\alpha$ of ${\bf H}_{\alpha}$, and therefore $N_W$, depends on the particular substrate. 

The ultimate objective of the reduction is then to find an effective Hamiltonian ${\bf H}_W'$ of dimension $N_W' \ll N_W$ whose eigenvalues and eigenvectors with large contribution of the TM orbitals are well reproduced. This implies to include the effect of the left-out states, which will be done using the perturbative Feshbach-Schur method.~\cite{Gustafson_Sigal_book_2011} Our strategy to reduce the dimensions of the problem is to diagonalize eq~\ref{eq:hmat} and choose a subset of eigenstates with a large weight on the original MLWFs of ${\bf H}_{sd}$. In order to further take into account the environment in the many-body configurations, we add states from the subspace of ${\bf H}_{\alpha}$.
To do this, we first diagonalize ${\bf H}_{\alpha}$, which leads to LCWO, and then choose the states of ${\bf H}_{\alpha}$ that have a large weight on the TM MLWFs. This becomes a second reduction that yields, in addition to the 6 MLWFs of the TM, a small set of $N_W'-6$ LCWO.

\subsubsection{Perturbative approach to the reduced crystal field Hamiltonian \label{PertFM}}
Let us introduce the eigenvalues and eigenvectors of the Wannier Hamiltonian,  ${ H}_W|M\rangle= \epsilon_M|M\rangle$.
The first $N_{sd}$ coefficients of the $|M\rangle$ vector give the weight of the vector in the TM-orbital subspace. In this way, we can choose the   $|M\rangle$ vectors that have a large component on each Wannier, $w$,
that has a $d$-character:
\begin{equation}
W_d(M)=\sum_{w} \left|\langle M|w\rangle\right|^2.
\label{eq:weight}
\end{equation} 
In order to find the reduced subspace of dimension $N_W'$, we need to include states from the environment corresponding to Hamiltonian ${\bf H}_{\alpha}$. The strategy is to take the states of the subspace of ${\bf H}_{\alpha}$ that have a large weight on the above $\{|M\rangle\}$ states with large TM component. These are the $|M\rangle$ states with $W_d(M)$ larger than a certain threshold $\Delta_{d}$.

For the description of the environment electronic structure, we use LCWOs, $|\beta\rangle$, obtained by diagonalizing ${\bf H}_{\alpha}$. Then, we will truncate the subspace given by ${\bf H}_{\alpha}$ by only using the $|\beta\rangle$ with a weight on $|M\rangle$ larger than a certain threshold, $\Delta_W$.

Hence, we can split the total Hilbert space in two: those states with a relevant $d$-orbital contribution, spanning a subspace of dimension $N_W'\equiv N(\Delta_d,\Delta_W)$, and the rest. The situation is then ideal to apply the perturbative 
Feshbach-Schur method.~\cite{Gustafson_Sigal_book_2011} This leads to the significantly reduced single-particle Hamiltonian, ${\bf H}_W'$, of dimension $N_W'$, that, in analogy to eq~\ref{eq:hmat}, can be expressed as:

\begin{equation}
{\bf H}_W'=
\left[
\begin{array}{cc}
 {\bf H}_{sd} &  {\bf V}_{sd,\alpha}'\\
 {\bf V}_{\alpha,sd}' &  {\bf H}_{\alpha}'\\
\end{array}
\right].
\end{equation}

\subsection{Multiplet Model  \label{Mult-methods}}

The previous section has determined the reduced one-body Hamiltonian, ${\bf H}_W'$, to be used in our multiplet calculations. In the present section we address the many-body aspects of a multiplet calculation.
The first point we address is the set up of electronic configurations leading to the crystal field. The second point is the correlation aspects induced by the Coulomb interaction.

The many-body basis set is built by considering $N_e$ electrons in $N_W'$ orbitals. The basis set is algorithmically built by considering all possible Slater determinants in order to ensure a complete many-body basis set in the reduced space of the multiplet. The Hamiltonian including the one-electron crystal- and ligand-field contributions, the spin-orbit and Zeeman interactions and the many-body electron-electron Coulomb repulsion is expressed as a matrix in this Slater-determinant basis set.

The crystal and ligand field contribution in ionic environments is frequently described in terms of point charges.\cite{Dagotto_book_2003} Although this approach does not provide a general good quantitative description, it is often used as a fitting procedure to electron spin resonance spectra (ESR),\cite{Abragam_Bleaney_book_1970} and we have successfully used it  to model the STM-ESR on a single Fe atom on MgO.\cite{ReinaGalvez2019}
Here, we take an alternative approach that goes beyond the point-charge description by including covalent bonding. Building on the above DFT-based calculations, we assume that the environment-related terms take the form
\begin{equation}
H_{\rm CF}=H_{\rm TM}+H_{\rm neigh}+H_{\rm hopp}
\label{hcf}
\end{equation}
where $H_{\rm TM}$ is the crystal field acting  only on the electrons on the TM orbitals,
\begin{equation}
H_{\rm TM}=
\sum_{w,w'} \langle w| {\bf H}_{sd}|w'\rangle 
  \sum_\sigma d_{w\sigma}^\dag d_{w'\sigma}
  +    \Delta_{s,d} \hat N_s.
\label{hcf_parts}
\end{equation}
Here, ${\bf H}_{\rm sd}$ is the Hamiltonian obtained in the previous section. The operators $d_{w\sigma}^\dag$ and $d_{w\sigma}$ denote the creation 
and annihilation of an electron with spin $\sigma$  in the MLWF $w$ of the TM with $d$-character.
The second term,  with $\hat N_s=\sum_{\sigma}d_{s,\sigma}^\dag d_{s,\sigma}$,  acts only on the MLWF with $s$-character of the TM. This second term
controls the relative occupation between the $s$ and $d$-shells. This term allows us to account for the different spatial extension of the $s$ and $d$-like MLWFs as well as to correct for the double-counting
included in the exchange-and-correlation potential of the DFT calculations.\cite{doublecounting}

The ligand field contribution $H_{\rm neigh}+H_{\rm hopp}$ contains both, an \textit{onsite} surface contribution, given by
$H_{\rm neigh}= \sum_{j} \langle j|{\bf H}_\alpha'|j\rangle  \sum_\sigma p_{j\sigma}^\dag p_{j\sigma}$, with $p_{j\sigma}^\dag$ ($p_{j\sigma}$) the creation 
(annihilation) operator of an electron with spin $\sigma$ in the $j$ LCWO of the surface, and the hopping between the surface orbitals and 
the adatom  states, $H_{\rm hopp}$,  given by
\begin{equation}
H_{\rm hopp}= \sum_{w,k,\sigma}\langle w|\textbf{V}_{sd,\alpha}'|k\rangle  d_{w\sigma}^\dag p_{k\sigma}+h.c.
\end{equation}
where $\textbf{V}_{sd,\alpha}'$ is the operator that couples the Wannier of the TM subspace with the LCWO of the reduced space given by ${\bf H}_\alpha'$.

We further approximate the space of configurations to only include configurations with a total number of active electrons on the surface, $\sum_{j,\sigma}\langle p_{j\sigma}^\dag p_{j\sigma}\rangle$, that are either 0 or 1. In other words, only electronic configurations $(sd)^{N_e}P^0$ and $(sd)^{N_e-1}P^1$ are considered (here $P$ denotes the surface orbitals). 

To evaluate the Coulomb electron-electron, spin-orbit and Zeeman matrix elements in the many-body basis set, we use hydrogenic atomic-like orbitals that agree with the MLWFs generators.\cite{Ferron2015,Ferron_Lado_prb_2015}
Since the surface orbitals are extended states associated to either the valence  or the conduction bands of the substrate, the corresponding Coulomb interaction is strongly reduced and can be safely neglected. Therefore, electron-electron interaction is accounted for only in the TM $s$ and $d$-atomic like orbitals, justifying our choice of basis set. Exact diagonalization of the Hamiltonian matrix in the many-body basis set ensures that exchange-and-correlation effects will be treated exactly within the subspace of interest.

\section{Results}
\begin{figure}[ht]
\includegraphics[width=0.5\textwidth]{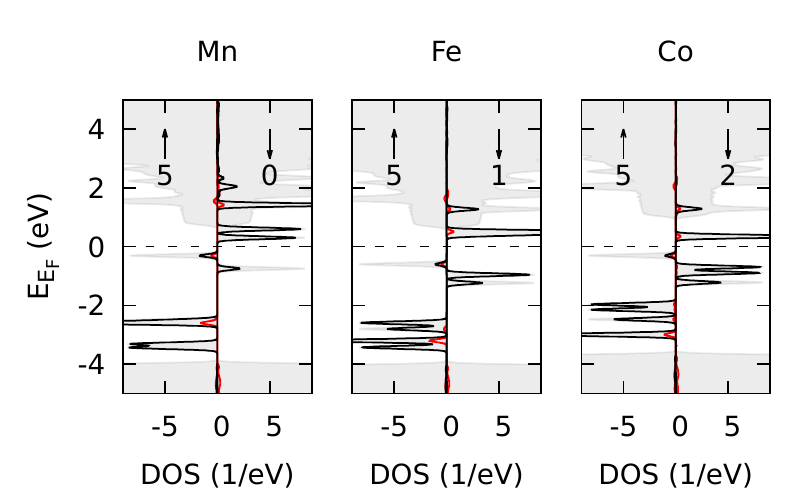}
\caption{DOS of the $3d$ states for Mn, Fe and Co (black lines) superimposed on the total DOS (grey). Arrows and numbers indicate the approximate occupation of up/down electrons as obtained from DFT. Spin-polarization on the TM adatom induces a small polarization on the underlying oxygen 2p orbitals (red).}
\label{fig:ldos}
\end{figure}
The DFT calculations were performed using the pseudopotential method and plane-waves as implemented in Quantum-Espresso.\cite{Giannozzi2017} Ultrasoft pseudopotentials from the PSL library\cite{DalCorso2014} were used for Fe, Co and MgO and a GBRV pseudopotenial\cite{Garrity2014a} was used for Mn. All pseudopotentials use the PBE parametrization for the exchange and correlation potential.\cite{Perdew1996b} Cutoffs for the expansion of the plane waves and charge density were chosen according to the SSSP pseudopotential verification database.\cite{Prandini2018} The bulk lattice constants for silver and MgO with PBE are a$_{\textrm{Ag}}$=4.16 \AA~and a$_{\textrm{MgO}}$=4.25~\AA, which results in a lattice mismatch of about 2\% (experimental value: 2.9\%).\cite{Wollschlager2001} To generate the slab structure we first used 10 monolayers (ML) of silver with the lateral lattice constant fixed to that of the PBE bulk silver and added up to 10 ML of MgO. The total energy of this system converges at a Monkhorst Pack grid of  $18\times 18\times 1$.\cite{Monkhorst1976} The system is padded by 10 \AA ~of vacuum in the z-direction and a dipole correction\cite{Bengtsson1999} was applied to decouple the slab from its periodic images. Grimme-d3 vdW correction\cite{Grimme2004} was used to get a accurate work function (4.32 eV),\cite{Prada2008} and inter-layer distances. 
Relaxing the system until the residual forces were less than $10^{-4}$ a.u.,\cite{Bieletzki2010, Malashevich2014} resulted in a distance of d$_\textrm{Ag-O}$=2.65 \AA~between the topmost MgO and the lowest Ag layers. We found that the first two layers of MgO led to a significant reduction of the work function of Ag, which stabilized at 3 ML (see Figure~\ref{fig:sup:WFAgMgO} in the supporting information), in agreement with previous studies of this system.\cite{Konig2009} Subsequently, we chose to use 3 ML of MgO as substrate for our model calculations and removed the silver. The adatom is placed on top of an oxygen site in a lateral $3 \times 3 \times 1$ MgO supercell and relaxed with the lowest layer of MgO frozen. The resulting adatom-oxygen distances in \AA ~are: 2.0 for Mn, 1.9 for Fe and 1.9 for Co. These values are in reasonable agreement with distances inferred from STM measurements and independent calculations.\cite{Baumann2015e, Willke2018} We note that for all adatoms we attempted to calculate the MAE by noncollinear total energy DFT calculations including SO  (details can be found in the supplementary information) and found that they were severely underestimated compared to the experiment by approximately a factor of 10 (see Table~\ref{tab:MAE_DFT}).

The resulting density of states (DOS) for each adatom as well as the underlying oxygen is shown in Figure~\ref{fig:ldos}. The DFT results correspond to the expected occupation of the 3d manifolds with 5, 6 and 7 electrons for Mn, Fe and Co, respectively. The electron transfer from the TM-4s state only accounts for ~0.2 electrons which does not allow us to assign a ``2+" charge state to the adatom based on DFT results. This can be either due to the difficulty of assigning an ``atomic charge" from DFT or due to on-site hybridization of 4s-3d on the adatom and inter-site 4s-2p with the underlying oxygen. We find that this ambiguity is best resolved by including the 4s manifold in the Wannier Hamiltonian and allowing for hopping between the 4s and 3d manifold whilst enforcing only an average occupation on the adatom.

\begin{figure*}
\centering
\includegraphics[width=0.9\textwidth]{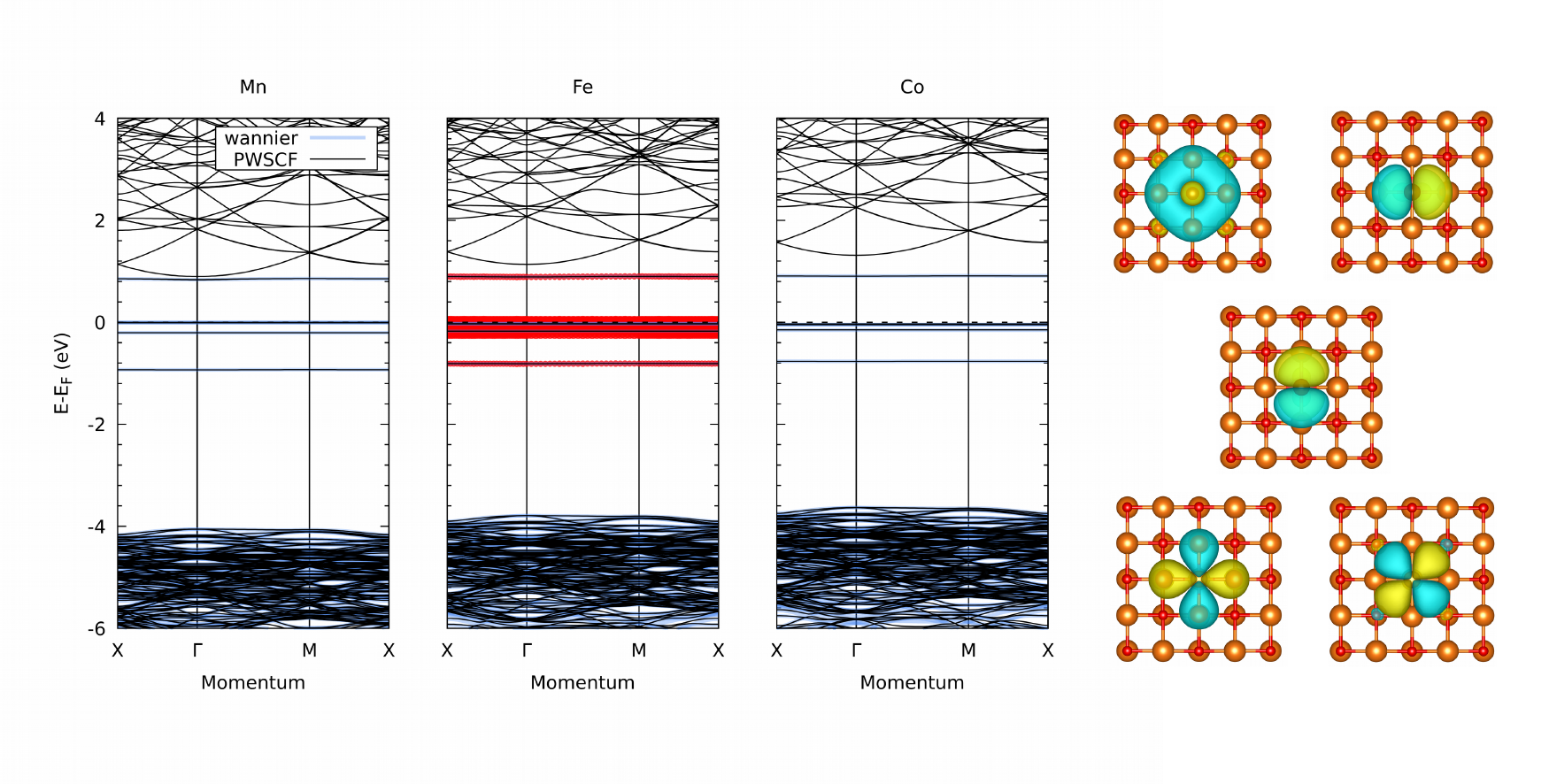}
\caption{(left) Results of the DFT calculation (black) superimposed with the Wannier interpolation (blue) following the wannierization procedure as described in the main text. (right) The resulting real-space plots of MLWFs corresponding to Fe bands with dominant $3d$ character (red) are shown. Their orbital character generally corresponds to the free atom case whilst their orientation reflects the underlying symmetry of the MgO substrate.}
\label{fig:mlwfs}
\end{figure*}

\subsection{Wannierization of the TM on MgO}

\begin{figure}[ht]
\begin{center}
\includegraphics[width=0.45\textwidth]{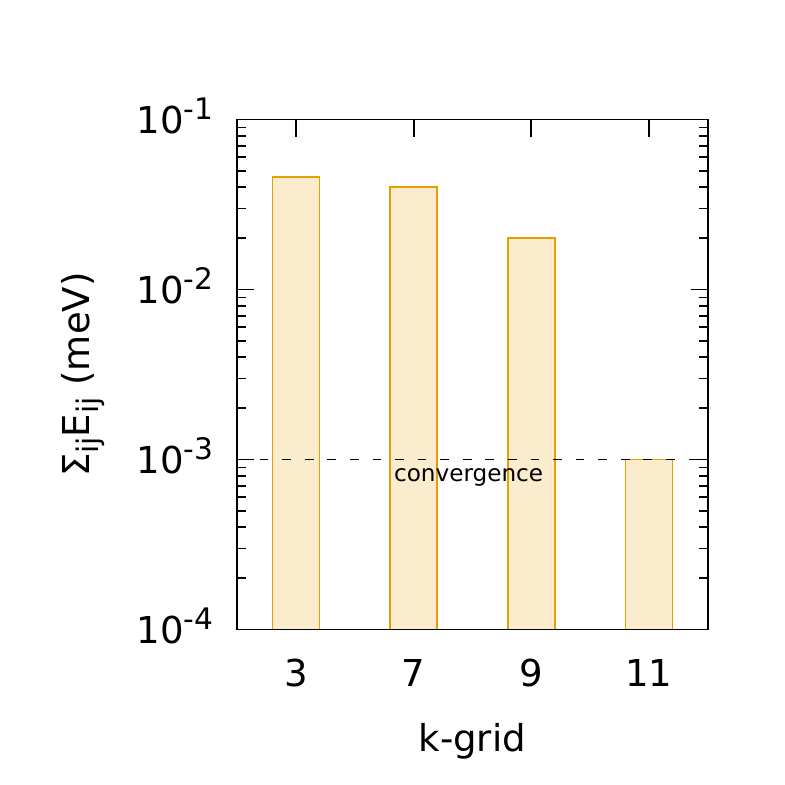}
\caption{Convergence of the sum of off-diagonal parts of the imaginary energy of the Wannier Hamiltonian ($\sum_{i,j}E_{i,j}, i\neq j$) for different $k$-grids. We consider the Wannier Hamiltonian converged when the sum of imaginary energies is $\approx 10^{-3}$ meV}
\label{convIM}
\end{center}
\end{figure}

{As criterion for localization of the MLWFs we consider the sum of the off-diagonal terms of the imaginary part of the Wannier Hamiltonian (at R=0), which should be strictly 0 in the case of true maximal localization.\cite{Brouder2007} We use an $11\times 11\times 1$ regular grid of k-points in the non spin-polarized calculation, corresponding to less than $10^{-3}$ meV for the sum of imaginary energies as shown in Figure~\ref{convIM}. Initial projections were defined on the 3d and 4s of the TM as well as oxygen 2p orbitals, which make up the valence bands of MgO, resulting in a total of $N_W=$168 MLWFs. These projections lead to small average spreads ($\Omega <1$ \AA$^2$) of the MLWFs. Further extending the Wannier basis set to include Mg did not lead to improvements.}

The results of the wannierization are shown in Figure~\ref{fig:mlwfs} for Mn, Fe and Co with the real-space plots of the MLWFs for the case of Fe:3d. In the following, we will focus on the case of Fe on MgO, but the discussion applies to Mn and Co as well. Real-space plots of the Wannier functions were created using VESTA.\cite{Momma2011}

\begin{table}[ht]
\caption{Value for the MAE from DFT total energy calculations compared to preferred magnetization axis and total barrier height from experiment (exp.). A negative sign of the MAE corresponds to out-of-plane magnetization
}
\label{tab:MAE_DFT}
 \begin{tabular}{|l| c ||c|  c|} 
 \hline
 TM &  MAE$^{\textrm{DFT}}$~(meV) &   MAE$^{\textrm{exp.}}$~(meV)\\  
 \hline\hline
Mn & -5.8  & (N.A.) \\ 
 \hline
Fe & -2.4 & -22 \cite{Baumann2014,Baumann_Donati_prl_2015}  \\
 \hline
Co  & -5.6 & -60\cite{Rau2014a} \\
 \hline
\end{tabular}
\end{table}

\begin{figure*}[ht]
\centering
\includegraphics[width=0.8\textwidth]{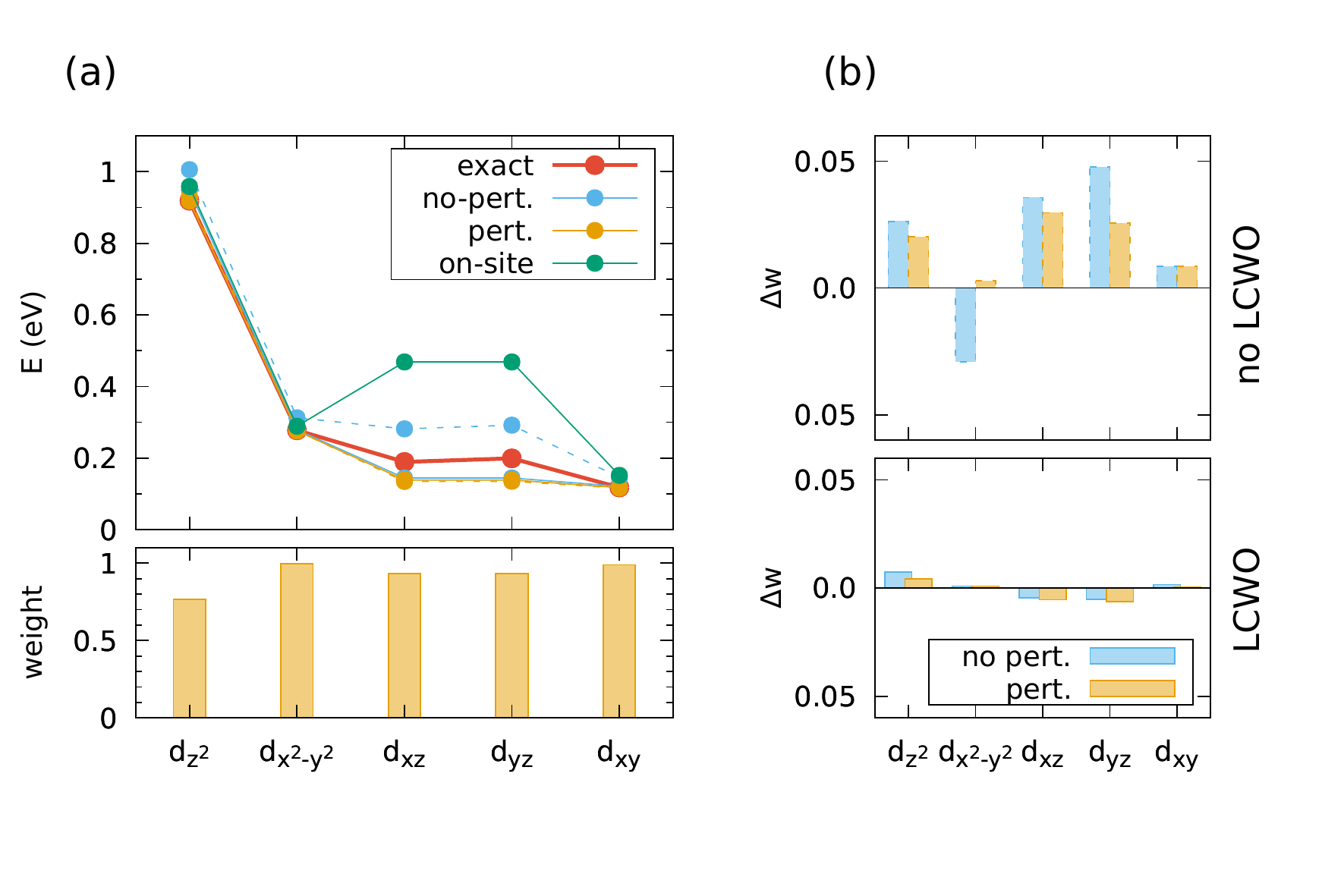}
\caption{(a) Single particle energy spectra for Fe on MgO for different levels of approximation. Shown are the on-site energies, the energy calculated using a non-perturbative approach (no pert.), the perturbative approach (pert.) and exact diagonalization of the Hamiltonian (exact). Solid lines refer to calculations using our LCWO approach, dashed lines correspond to calculations without LCWO. The states are labeled according to the dominant contribution of the wave functions with the respective weights shown. (b) relative weights of the wave functions calculated as difference to the weights from exact diagonalization for non-perturbative and perturbative approach without and with LCWO.}
\label{fig:comp_LCWO}
\end{figure*}

As mentioned before, the number of MLWFs used to reproduce the DFT energy bands is too large to create a manageable basis set to span the Hilbert space of the many body problem, so LCWO and the perturbative Feshbach-Schur treatments are used to speedup the convergence and reduce the computational cost.
The basis reduction has been carried out taking $\Delta_d=0.02$ and $\Delta_W=0.003$, which leads to a dimension $N_W'=26$. The results of the combination of LCWO with the Feshbach-Schur method are then compared with the spectrum of the whole Wannier Hamiltonian, see Figure \ref{fig:comp_LCWO}. In addition, they are compared with the spectra without LCWO and neglecting the perturbative effect of the states left out of the reduced subspace. An analysis of the data in Figure~\ref{fig:comp_LCWO} clearly shows that the perturbed solution already provides a considerable improvement in energy convergence, which is further enhanced by the LCWO. In the case of very efficient decoupling layers,improvements seem moderate but LCWO leads to a significant improvement in the case of more strongly hybridized substrate-adatoms, as it happens on CuCl(111) or Cu$_2$N/Cu(100) surfaces.

\subsubsection{Multiplet of Fe on MgO}
\begin{figure*}[t]
\centering
\includegraphics[width=1.0\textwidth]{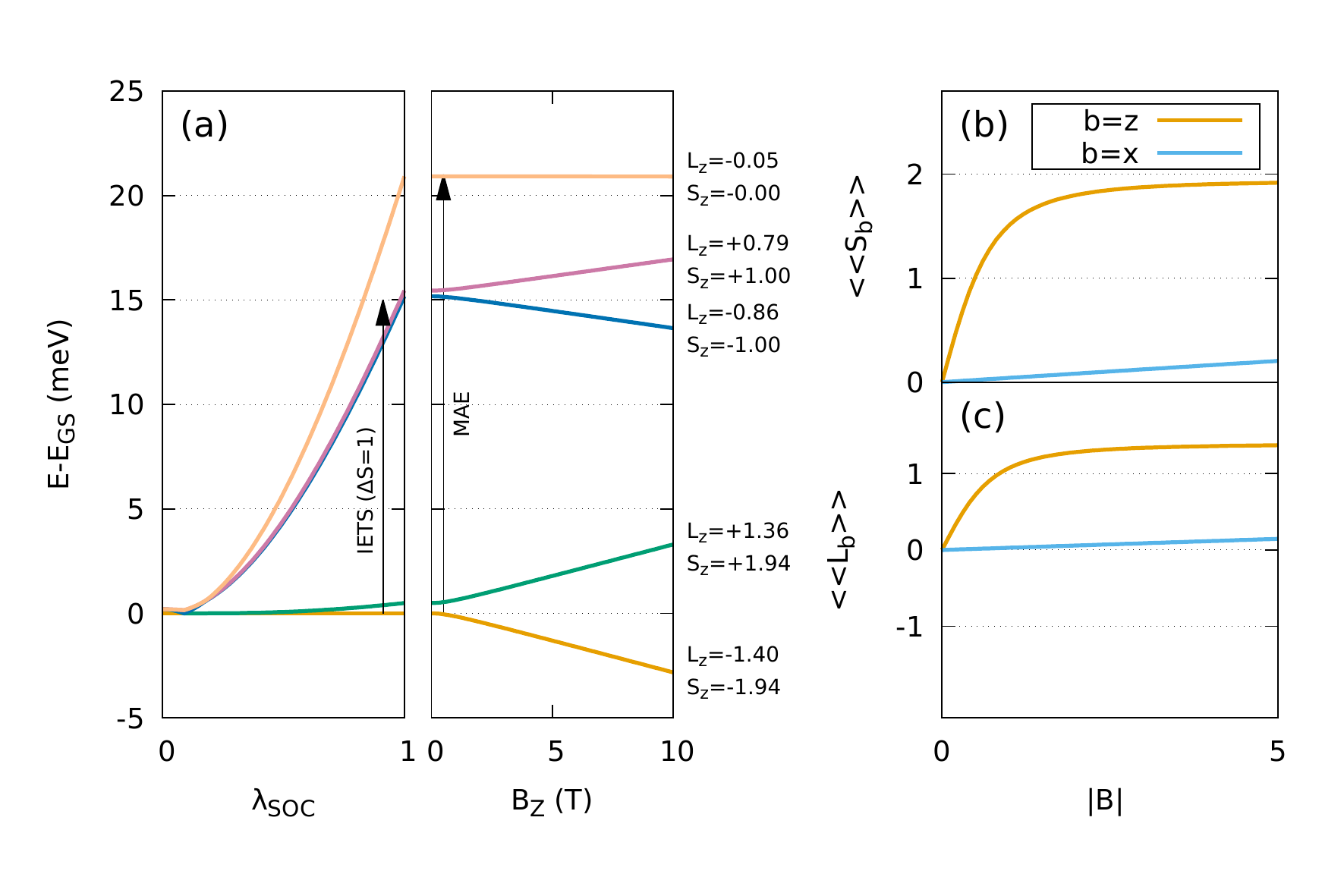}
\caption{ (a) Low-energy spectrum of Fe on MgO. The energy relative to the ground-state is shown as function of increasing spin-orbit ($\lambda_{SOC}$) coupling and external out-of plane magnetic field strength $B_z$. Labels indicate the expectation values for $\langle S_z \rangle$ and $\langle L_z \rangle$ at $B=5$ T. The IETS-ZFS and MAE are shown schematically. The splitting of the $S_z=\pm 2$ states at zero field is most likely overestimated in our calculation (see main text for details). (b) and (c) show the thermal average (at T=0.5 K)  for the spin and orbital angular momentum (in units of $\hbar$) measured in-plane ($b=x$) and out-of-plane ($b=z$) for a magnetic field applied in the same direction. The saturation with $b=z$ indicates the out-of-plane (easy-axis) preferred magnetization.}
\label{fig:multipletFe}
\end{figure*}

The case of Fe on MgO has been studied extensively experimentally and is therefore an excellent case to benchmark our calculations. Inelastic electron tunneling spectroscopy (IETS) and x-ray magnetic circular dichroism (XMCD) have shown that this system has preferred out-of-plane magnetization with a splitting between ground-state and first excited state of 14 meV at zero external magnetic field, corresponding to the zero-field split (ZFS) measured by IETS (selection rule $\Delta S=0,\pm1)$.\cite{Baumann_Donati_prl_2015,Baumann_Paul_science_2015} The orbital moment is unquenched and large which leads to an effective g-factor of 2.6, significantly larger than the g-factor for a free electron of 2.0.\cite{Baumann_Paul_science_2015,Baumann_Donati_prl_2015}\\
We perform multiplet calculations on a reduced basis including 6 orbitals on the adatom and 20 surface orbitals ($N_W=26$). The Coulomb interaction is parametrized by an average repulsion $U_{\rm eff}=U-J=5$ eV. We consider $N_e=9$ electrons in the phase space with a final electron occupation of $4s^2 3d^6$ for the adatom and one remaining electron on the surface. The results of our calculation are shown in Figure \ref{fig:multipletFe}. The calculated spectrum is in excellent agreement with the experimental findings. The IETS ZFS of 14 meV, the MAE of 21 meV, as well as the unquenched orbital moment $L_z$ are on top of the experimental values.\cite{Baumann2014} The unquenched orbital moment leads to an effective $g$-factor of 2.56 in our calculation. The splitting between the $S_z=\pm 2$ states implies the presence of a small quantum spin tunneling (QST), which leads to a null average magnetization at zero field. The magnitude of the QST only slowly decreases with the basis size which is why the calculated value exceeds the experimentally found value of approximately 1 neV.\cite{Paul_Yang_natphys_2017} We also calculated the expectation values for $\langle S_z \rangle$ and $\langle L_z \rangle$ (details on the calculation of observables from the multiplet model can be found in the appendix) for in-plane and out-of-plane external magnetic field. As shown in Figure~\ref{fig:multipletFe} (b) and (c) they quickly saturate for out-of-plane (indicated by b=z) magnetic field which confirms the preferred out-of-plane magnetization axis.

\begin{figure}
\includegraphics[width=0.5\textwidth]{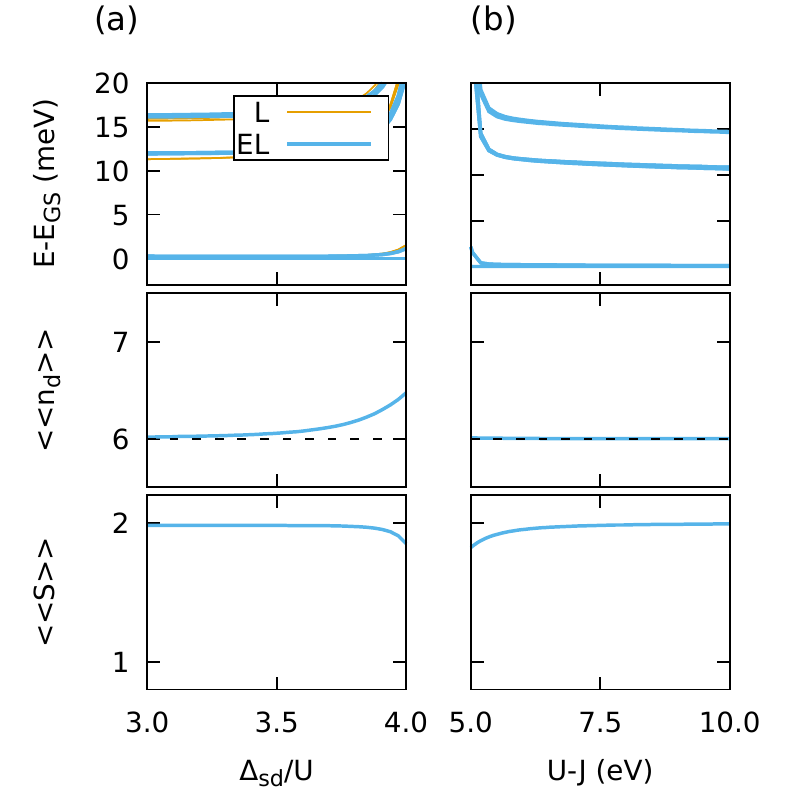}
\caption{(a) Dependence of multiplet energies $\textrm{E}$-E$_{\textrm{GS}}$ (for two different bases), occupation of the d-manifold $<<\textrm{n}_{d}>>$ and total spin $<<\textrm{S}>>$ of the adatom as function of the adatom $s-d$ energy splitting $\Delta_{sd}$ and (b) their dependence on the effective U-J parameter.}
\label{fig:delta_SD}
\end{figure}

The explored configuration has $\langle\langle N_d\rangle\rangle=6$ with a total spin $S=2$, as found experimentally,\cite{Baumann_Donati_prl_2015} and  the the best agreement was found for the 4s$^2$3d$^6$ configuration. To study the robustness of the method we explored the effects of variations of the relative splitting between $s$ and $d$ orbitals $\Delta_{sd}$ on the adatom, as well as the value of the effective Hubbard $U_{\textrm{eff}}$ as introduced in the treatment of the crystal field, eqs \ref{hcf_parts} and \ref{hcoul}. This is shown in Figure~\ref{fig:delta_SD}. As observed, appreciable changes in the MAE only appears when the occupation of the TM orbitals changes. Additional analysis (not shown) of the dependence on $E_0$ demonstrates that, for $U_{\rm eff}\le E_0\le 6U_{\rm eff}$, variations of the ZFS are below 0.1\%.
The 4s$^2$3d$^6$ configuration is maintained as long as $\Delta_{sd}\leq 4~ U_{\rm eff}$. The multiplet solution is robust in a wide range of $5\; {\rm eV}\leq U_{\textrm{eff}}\leq 10$ eV, which allows us to assign a relatively conservative value of $U_{\textrm{eff}}=5$ eV. This value agrees well with calculated values for $U$ using linear response, which in our case gave $U_{\textrm{eff}}=5.2$ eV.\cite{Cococcioni2005, Linscott2018}

\subsubsection{Multiplet of Co on MgO}
\begin{figure*}
\centering
\includegraphics[width=0.9\textwidth]{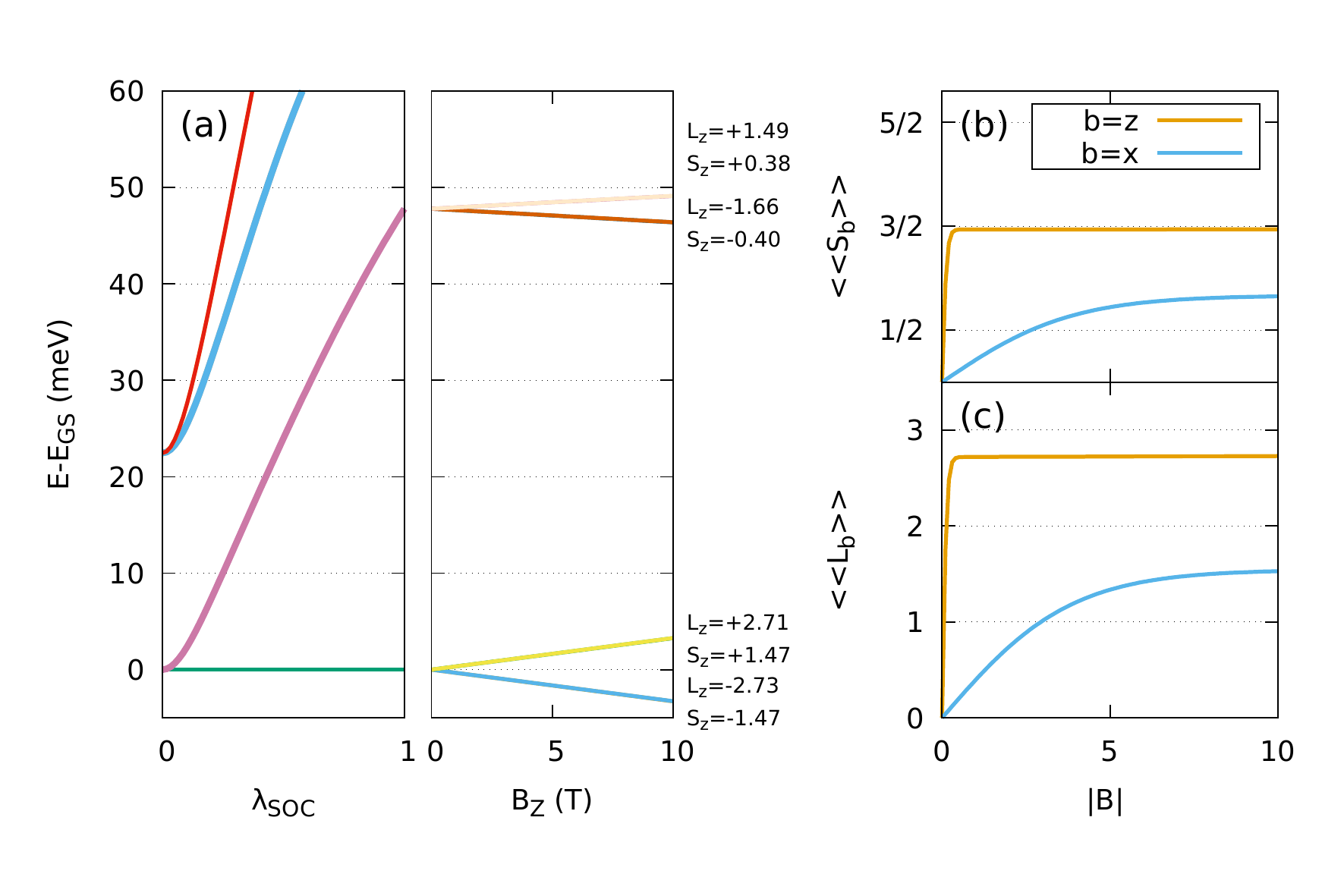}
\caption{ (a) Low-energy spectrum of Co on MgO. The energy relative to the ground-state is shown as function of increasing spin-orbit ($\lambda_{SOC}$) coupling and external out-of plane magnetic field strength $B_z$. Labels indicate the expectation values for $\langle S_Z \rangle$ and $\langle S_L \rangle$ at $B=5$ T. (b) and (c) show the thermal average (at T=0.5 K)  for the spin and orbital angular momentum (in units of $\hbar$) measured in-plane ($b=x$) and out-of-plane ($b=z$) for a magnetic field applied in the same direction. The saturation with $b=z$ indicates the out-of-plane (easy-axis) preferred magnetization. }
\label{fig:multipletCo}
\end{figure*}

It has been experimentally confirmed that Co on MgO exhibits the largest possible anisotropy of any TM on MgO, due in part to large contributions from the unquenched orbital moment with $L_z=3$, leading to a value of 57 meV for the ZFS measured by IETS with an out-of-plane easy axis.\cite{Rau2014a} Both observations are confirmed by our calculations indicating that Co has an $S_z=3/2$ ground state with preferred out-of plane orientation of the magnetization and a ZFS of about 48 meV which only slightly underestimates the experimental value. The results are shown in Figure~\ref{fig:multipletCo}.

\subsubsection{Multiplet of Mn on MgO}
\begin{figure*}
\centering
\includegraphics[width=0.9\textwidth]{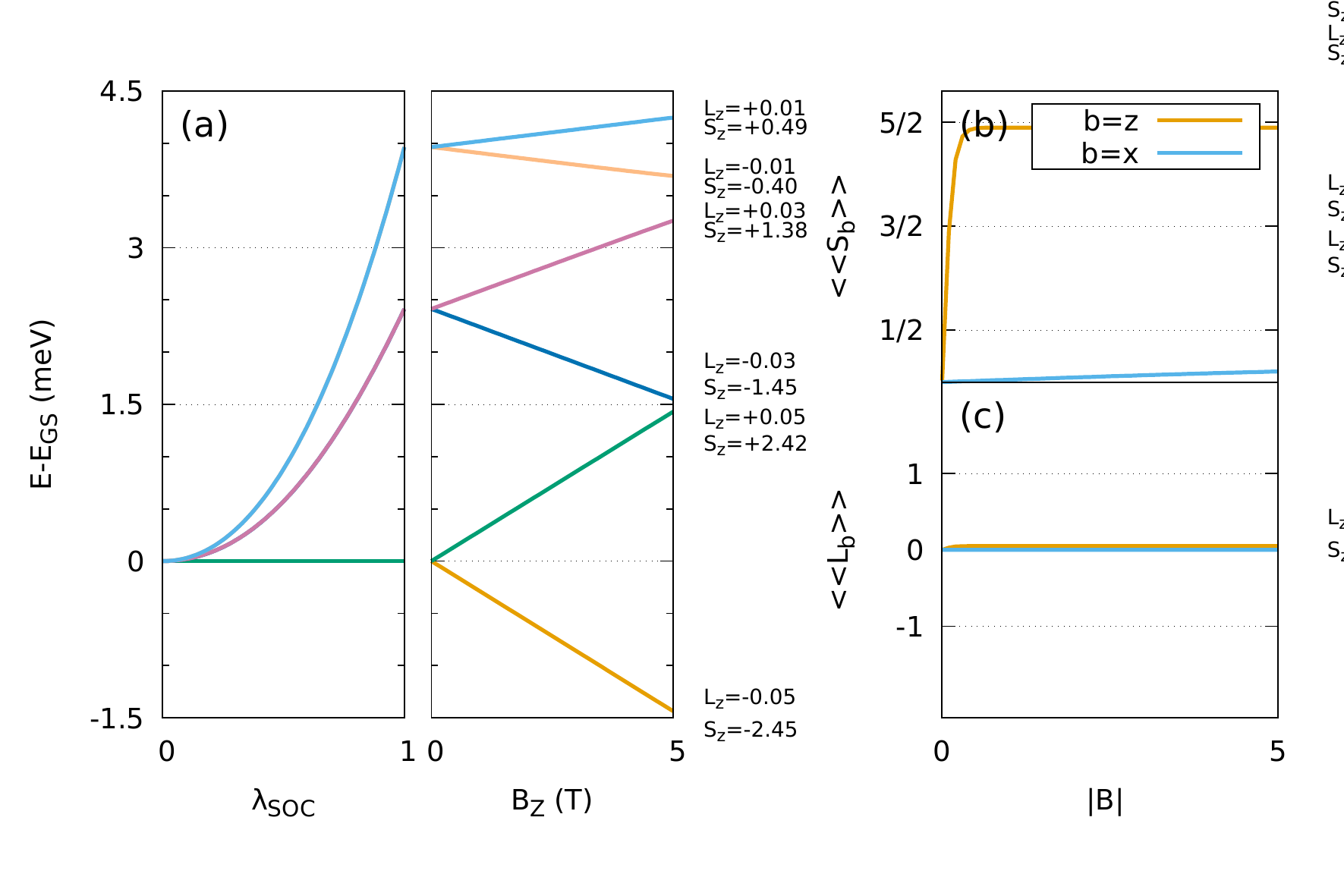}
\caption{(a) Low-energy spectrum of Mn on MgO. The energy relative to the ground-state is shown as function of increasing spin-orbit ($\lambda_{SOC}$) coupling and external out-of plane magnetic field strength $B_z$. Labels indicate the expectation values for $\langle S_Z \rangle$ and $\langle S_L \rangle$ at 5 T, respectively. (b) and (c) show the thermal average (at T=0.5 K)  for the spin and orbital angular momentum (in units of $\hbar$) measured in-plane ($b=x$) and out-of-plane ($b=z$) for a magnetic field applied in the same direction. The saturation with $b=z$ indicates the out-of-plane (easy-axis) preferred magnetization.}
\label{fig:multipletMn}
\end{figure*}

Spin transitions of Mn on MgO have not been observed so far. This might be surprising, as previous experiments have shown that Mn shows spin transitions on Cu$_2$N, with a very small anisotropy visible in the IETS spectrum at around 0.1 meV, interpreted as transition from the $|m_s=5/2\rangle\rightarrow |m_s=3/2\rangle$ state.\cite{Loth2010,Hirjibehedin2007} The expected low splitting poses an experimental challenge, but it is further complicated by the observation that Mn on MgO can absorb on top of oxygen or on an O-O bridge site exhibiting different spectral features in IETS without clear spin transition.\cite{Baumann2015e} Here, we will only consider the ``oxygen top'' configuration which has the lowest energy in our DFT calculation (a comparison of absorption energies is given in Table \ref{tab:Mn-abs-energy}), comparable with the scenario of  
 Fe and Co. The multiplet spectrum for Mn is shown in Figure~\ref{fig:multipletMn}. The preferred magnetization axis lies out of plane, and the ground-state has $|S_z|=5/2$ with strongly quenched orbital moment due to the half-filling of the 3d shell with a resulting ZFS of 2.5 meV.
The resulting ZFS of 2.5 meV is small compared to Co and Fe, following the general trend of strongly reduced anisotropy when replacing Co or Fe by Mn on a Cu$_2$N substrate.\cite{Hirjibehedin2007,Otte_Ternes_natphys_2008}

\section{Conclusions}
Here, we have made use of an ab-initio based electronic multiplet calculation to study the magnetic properties of 3$d$ transition metal atoms on a MgO surface. The method combines the benefits of a linear combination of atomic orbitals, which efficiently accounts for the symmetry of the crystal and ligand field, and the low spread and quasi-atomic character of maximally localized Wannier functions. By applying the method to Mn, Fe and Co on MgO we can accurately reproduce experimental findings, in particular, the preferred out-of-plane magnetization of all adatoms, the magnetic ground state and spin and orbital moments of the states involved in the transitions observed in IETS experiments. For Fe and Co, where experimental data are available, we obtain 14 meV and 48 meV for the lowest IETS transitions, which agrees well with 14 meV and 58 meV obtained by experiments. For Mn on MgO we can predict a spin excitation at 2.5 meV which so far has not been observed. 
In addition, we have analyzed the robustness of our approach with the parameters that can not be precisely known from a first-principle calculation, such as the average onsite repulsion energy, $U$, or the actual atomic charge. We have shown that, far from critical regions where charge transfer occurs, our method provide stable solutions. We believe that our method greatly enhances our ability to systematically search for substrate-magnetic adatom combinations with high magnetic anisotropy energies, which are promising candidates for magnetic data storage at the atomic limit.

\begin{table*}[ht]
\caption{Preferred orientation of the magnetization axis, $\Delta E_{0\rightarrow 1}$ for a spin excitation $|0\rangle\rightarrow |1\rangle$, $S_z$ of the ground state, effective g$^*$-factor for Mn, Fe and Co on MgO. Values in parenthesis are experimental reported
values.}
\label{tab:summary}
 \begin{tabular}{|l|| c| |c|  c| c|} 
 \hline
 TM Species &  axis & $|S_z|$ &  $\Delta E_{0\rightarrow 1}$ (meV) & g$^*$-factor \\  
 \hline\hline
Mn & out-of-plane & $\frac{5}{2}$ & 2.5 & 1.96 (1.90 on CuN)\cite{Hirjibehedin2007} \\ 
 \hline
Fe & out-of-plane & $2$ & 15 (14)\cite{Baumann2014,Baumann_Donati_prl_2015} & 2.60 (2.57)\cite{Baumann_Donati_prl_2015} \\
 \hline
Co  &  out-of-plane & $\frac{3}{2}$ &48 (58)\cite{Rau2014a},  & 3.79 (3.7)\cite{Rau2014a}  \\
 \hline
\end{tabular}
\end{table*}

\begin{acknowledgement}
The authors are grateful for stimulating discussions with A.J. Heinrich and T. Choi. C.W. acknowledges funding from the Korean Institute of Basic Science under IBS-R027-D1.
FD acknowledges financial support from Basque Government, grant IT986-16 and Canary Islands program {\it Viera y Clavijo}  (Ref. 2017/0000231). N.L. and J.R.G acknowledge funding from the Ministerio de Ciencia e Innovaci\'on Grant No. RTI2018-097895-B-C44 and FEDER funds.
\end{acknowledgement}

\section{Appendix}\label{app:param_MO}
\subsection{Expressions of the multiplet Hamiltonian in second quantization \label{appendixHm}}
The Coulomb term $H_{{\rm Coul}}$ can be written as (eq~\ref{hcoul})
\begin{equation}
\begin{split}
H_{{\rm Coul}} & =\frac{1}{2}\sum_{m,m'\atop n,n'}
V_{mnm'n'}
\sum_{\sigma\sigma'}d_{m\sigma}^\dag d_{n\sigma'}^\dag d_{n'\sigma'}d_{m'\sigma}\\
 &- E_0 \hat N_{\rm TM}.
 \end{split}
\label{hcoul}
\end{equation}
The TM orbitals, denoted by  $\phi_{m}(\vec{r})$,  are assumed to be equal to the product of a radial hydrogenic function (with  effective charge $Z$ and a  effective Bohr radius $a_\mu$) and a spherical harmonic. In the case of hydrogenic wavefunctions, the Coulomb integrals $V_{mn,m'n'}$ can be calculated analytically in terms of the Wigner 3-$j$ symbols and the Slater integrals $F^n(3d)$ and $F^n(4s)$.~\cite{Slater_Philips_pt_1974} The strength of the interaction can be defined in terms of a single repulsion parameter,~\cite{Ferron2015} the average onsite repulsion energy $U=\langle V_{nn,nn}\rangle$. 

 The last term in eq (\ref{hcoul}), where $\hat N_{\rm TM}=\sum_{m,\sigma} d_{m\sigma}^\dag d_{m\sigma}$ and $E_0$ is an onsite energy, controls the occupation of the magnetic atom, which can fluctuate in one unit due to the hoppings to the surface orbitals.

The spin-orbit coupling $H_{SO}$  is assumed to be non-zero only on the TM orbitals. It reads as 
\begin{equation}
H_{\rm SO}=\zeta\,\sum_{mm',\sigma\sigma'} \langle m\sigma|\vec \ell\cdot \vec S|m'\sigma'\rangle
d_{m\sigma}^\dag d_{m'\sigma'},
\end{equation}
 where $\zeta$ is the single particle spin-orbit coupling of the 
$d$-electrons. This term is frequently expressed as
$\lambda \vec L\cdot \vec S$ with $\vec L$ and $\vec S$ the total orbital and spin angular momentum, where $\lambda=\pm \xi/2S$ with the plus (minus) sign for half-filled or below (above)~\cite{Abragam_Bleaney_book_1970}. 

The Zeeman term contains two contributions: the Zeeman interaction of the TM electrons, $H_{\rm Ze}^{\rm TM}$, and the Zeeman contribution due to the unpaired surface electrons, $H_{\rm Ze}^{\rm neigh}$. The former is given by
\begin{equation}
\label{zeemMA}
H_{\rm Ze}^{\rm TM}=\mu_B \vec B\cdot \sum_{mm',\sigma\sigma'} \langle m,\sigma |\left(\vec l+g\vec S\right)
|m'\sigma'\rangle d_{m\sigma}^\dag d_{m'\sigma'},
\end{equation}
where $g=2$ and $\mu_B$ is the Bohr magneton.
For the surface electrons, we assume Pauli's diamagnetic response,~\cite{Kittel_Fong_book_1987} with an induced spin magnetic moment $m_P(B)\approx \rho(\epsilon_F)\mu_B^2 B$. Hence, 
we have that $H_{\rm Ze}^{\rm neigh}=2m_{p}(B) \vec B\cdot \sum_{\sigma\sigma'} \langle \sigma |\vec S
|\sigma'\rangle \sum_j p_{j\sigma}^\dag p_{j\sigma'}$.

\subsection{Observables from the multiplet model}
The solution of Hamiltonian (\ref{eq:htot}) leads to a set of eigenvalues $E_M$ and eigenvectors $|M\rangle$ of the many body problem. With this information we can look at the expectation values of several observables and their thermal equilibrium mean values.
In particular, we can be interested in evaluating the average total spin and orbital angular momentum $S$ and $L$ respectively, or their components along a given direction $a$, $S_a$ and $L_a$. 
The expectation values of an operator $\hat O$ will be denoted by $\langle O\rangle_M\equiv \langle M|\hat O|M\rangle$, while the thermal equilibrium mean values will be denoted by $\langle\langle \hat{O}\rangle\rangle={\rm Tr}[\hat\rho\hat O]$, with $\hat \rho$ the density matrix corresponding to thermal equilibrium.

 In first place, we will be interested in the average occupation of the $s$ and $d$ shells of the TM, $\langle \langle N_s\rangle\rangle$ and $\langle \langle N_d\rangle\rangle$ respectively, together with the average occupation of the surface orbitals, which we will denote as $\langle \langle N_{P_S}\rangle\rangle$. In addition, the average spin $\langle \langle \vec S\rangle\rangle$ and orbitals moments $\langle \langle \vec L\rangle\rangle$ will be defined in the same way.

\begin{suppinfo}
The following files are available free of charge.
\begin{itemize}
  \item Additional calculations regarding the influence of the electric field on the electronic states of the adatom, and calculation of MAE from DFT
\end{itemize}
\end{suppinfo}

\providecommand{\latin}[1]{#1}
\makeatletter
\providecommand{\doi}
  {\begingroup\let\do\@makeother\dospecials
  \catcode`\{=1 \catcode`\}=2 \doi@aux}
\providecommand{\doi@aux}[1]{\endgroup\texttt{#1}}
\makeatother
\providecommand*\mcitethebibliography{\thebibliography}
\csname @ifundefined\endcsname{endmcitethebibliography}
  {\let\endmcitethebibliography\endthebibliography}{}


\pagebreak
\begin{center}
\textbf{\large Supplemental Material}
\end{center}
\setcounter{equation}{0}
\setcounter{figure}{0}
\setcounter{table}{0}
\setcounter{section}{0}
\setcounter{page}{1}
\makeatletter
\renewcommand{\theequation}{S\arabic{equation}}
\renewcommand{\thesection}{S\arabic{section}}
\renewcommand{\thefigure}{S\arabic{figure}}
\renewcommand{\thetable}{S\arabic{table}}
\renewcommand{\bibnumfmt}[1]{[S#1]}
\renewcommand{\citenumfont}[1]{S#1}
\section{Absorption energy differences for Mn on MgO}
It has been shown that Mn does not only absorb on the ``oxygen top'' site but also on bridge sites. We find that oxygen top is the lowest in energy with all other absorption sites about 0.5 eV less favorable in energy when Mn is assumed to be in its neutral state. This supports previous reports that have indicated that Mn on bridge sites might have be in a different charge state.\cite{Baumann2014}

\begin{table}[ht]
\caption{Absorption energy difference in meV for Mn on different sites relative to the oxygen top absorption site. The total magnetic moment \text{M} per cell for each configuration is given}
\label{tab:Mn-abs-energy}
 \begin{tabular}{|l|| c| c|} 
 \hline
 Absorption site & $\Delta$E (meV) & \text{M} ($\mu_B$) \\  
 \hline\hline
Oxygen-top & 0.0 & 5.0  \\ 
 \hline
Mg-top & 459 & 5.0  \\
 \hline
Mg-Mg bridge & 442 & 5.0  \\
 \hline
Mg-O bridge & 417 & 5.0  \\
 \hline
\end{tabular}
\end{table}

\section{MAE from noncollinear total energy calculations}\label{sec:anisotropy}

To demonstrate the failure of noncollinear SO calculations to predict magnetic anisotropy energies it is helpful to map the DFT calculations to an anisotropy spin Hamiltonian (eq~\ref{eq:anisotropy_spin})
\begin{equation}
\label{eq:anisotropy_spin}
\textbf{H}=DS_z^2+E\left(S_x^2-S_y^2\right),
\end{equation}
with only the axial anisotropy parameter D$\neq0$ and the transverse anisotropy parameter E=0 (the latter being justified by the isotropic environment in $x$ and $y$-direction on the MgO substrate).
 
The total DFT energy for spin orientation perpendicular (E$_{\perp}$) and parallel (E$_{\parallel}$) correspond to the evaluation of the DFT Hamiltonian with spin pointing in $z$ and $x$ direction and are therefore related to D as: 

\begin{equation}
\label{eq:ansiotropy_D}
D=\frac{2\left(E_{\perp}-E_{\parallel}\right)}{S\left(2S-1\right)}.
\end{equation}

Equation~\ref{eq:ansiotropy_D} allows us to calculate the total barrier height as difference of eq~\ref{eq:anisotropy_spin} evaluated for the spin ground state and the highest excited state, e.g. for Fe with a ground state of $S_z=\pm2$ and the excited state being $S_z=0$, the anisotropy energy takes the form $\textrm{MAE}^{\textrm{DFT}}=|D|(2^2-0)=4|D|$.
Noncollinear SO calculations were carried out using fully relativistic pseudopotentials from the PSL library.\cite{DalCorso2014}

\section{Work-function of the MgO capped Ag(100)}
The work function (WF) was calculated as the difference between the slab Fermi level and vacuum level, where the electronic potential becomes flat. Previous studies have put the WF of Ag (100) at $\Phi=4.2-4.3$ eV, which is in good agreement with our result of 4.24 eV.\cite{Prada2008} 
\begin{figure}
\centering
\includegraphics[width=1.\linewidth]{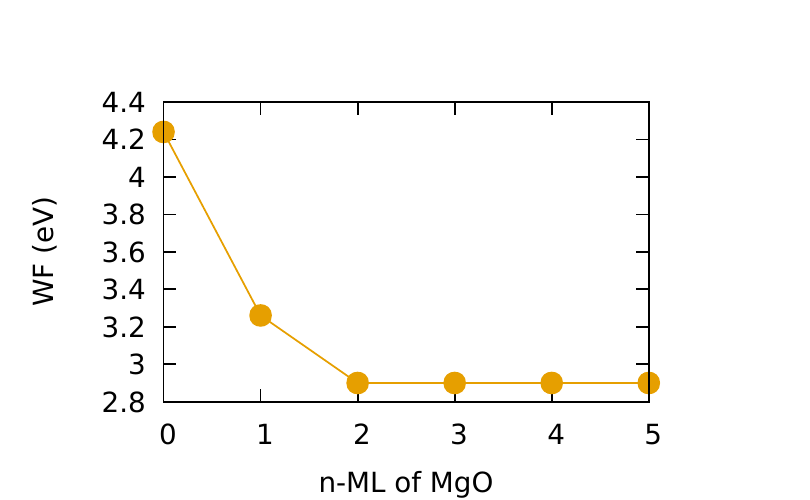}
\caption{Modification of the Ag(100) work function due to n=1..6 ML of MgO. The first two MgO layers strongly suppress the work function after which it saturates.}
\label{fig:sup:WFAgMgO}
\end{figure}

\begin{figure}
\centering
\includegraphics[width=1.\linewidth]{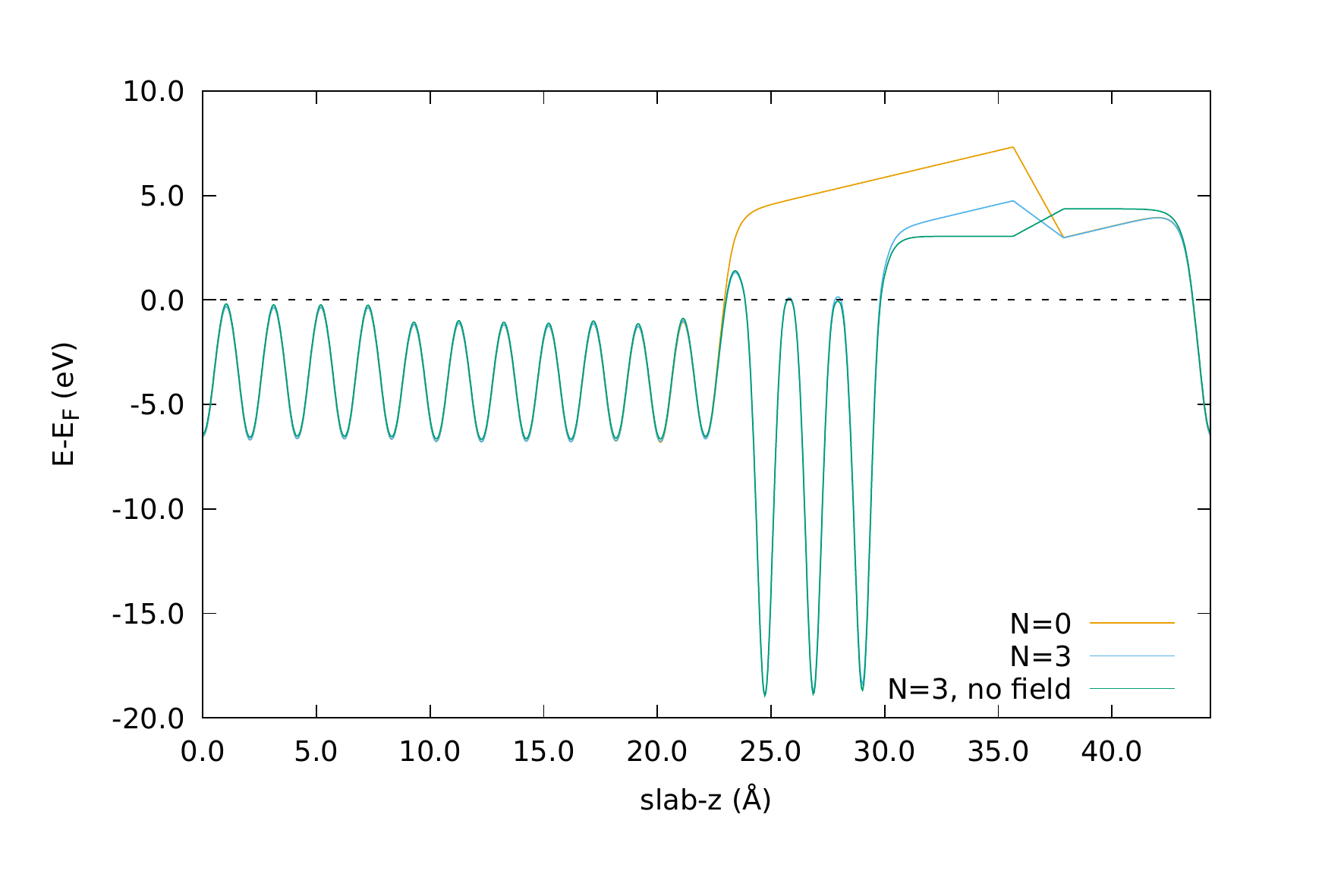}
\caption{In-plane averaged Kohn-Sham potential relative to the Fermi level for the Ag(100)/MgO slab. The cases of bare silver (N=0), capped silver (N=3) are shown with and without applied field (V=2.5 V/nm).}
\label{fig:sup:e_field}
\end{figure}

We further find that 3 ML of MgO perfectly screen the underlying silver substrate from fields between STM tip and Ag substrate. In order to illustrate this, we have applied a homogeneous electric field ($V\leq$5 V/nm) across the slab and find that the surface Ag potential does not shift from the case without a field when capped with MgO, as shown in Figure~\ref{fig:sup:e_field}.

\begin{figure}
\centering
\includegraphics[width=1.\linewidth]{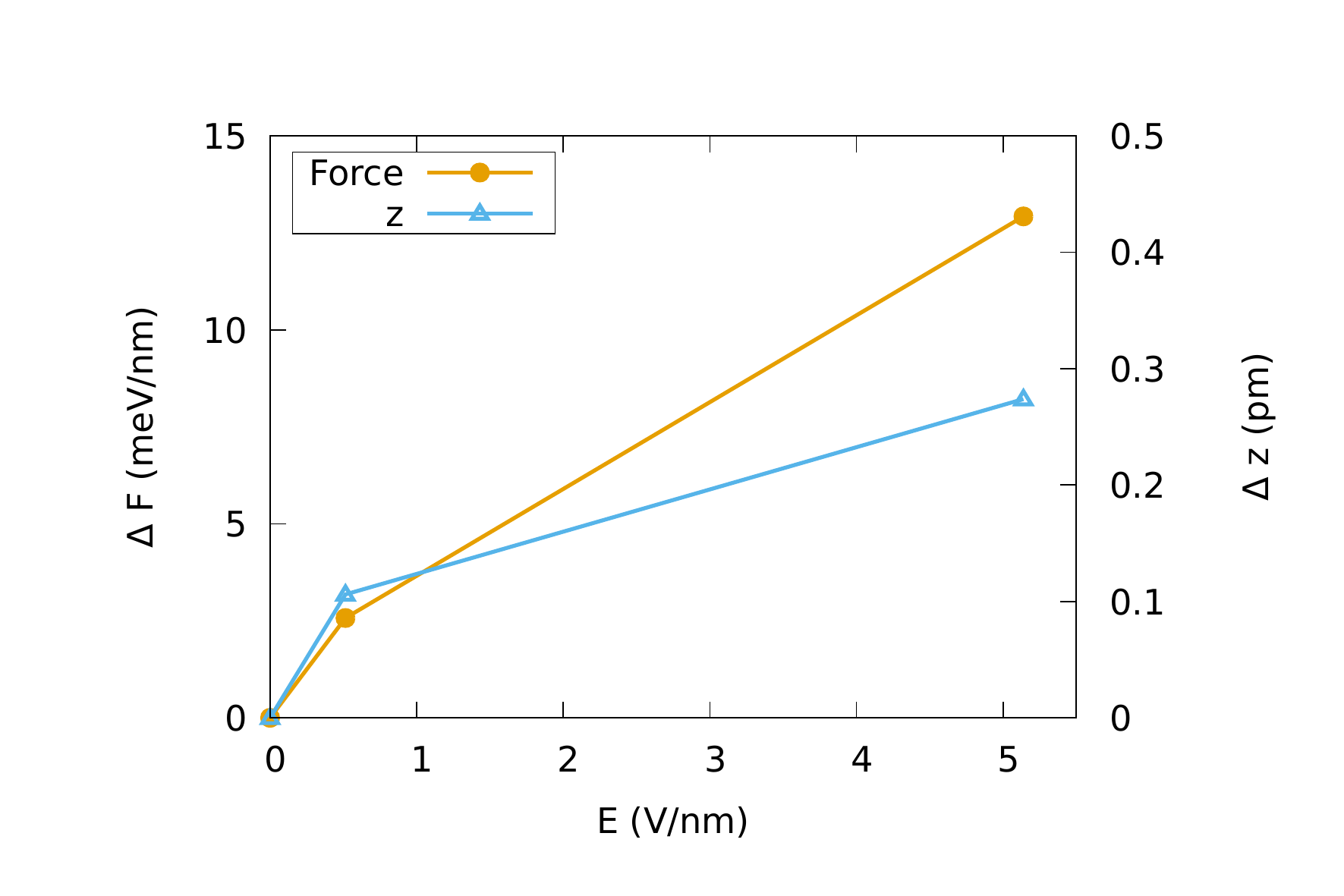}
\caption{Effect of the electric field strength on the adatom. The resulting force is small and relaxation of the adatom under field leads to a total displacement of less than 0.3 pm at an applied field of 5 V/nm. }
\label{fig:sup:force_Fe}
\end{figure}

\begin{figure*}
\centering
\includegraphics[width=0.9\textwidth]{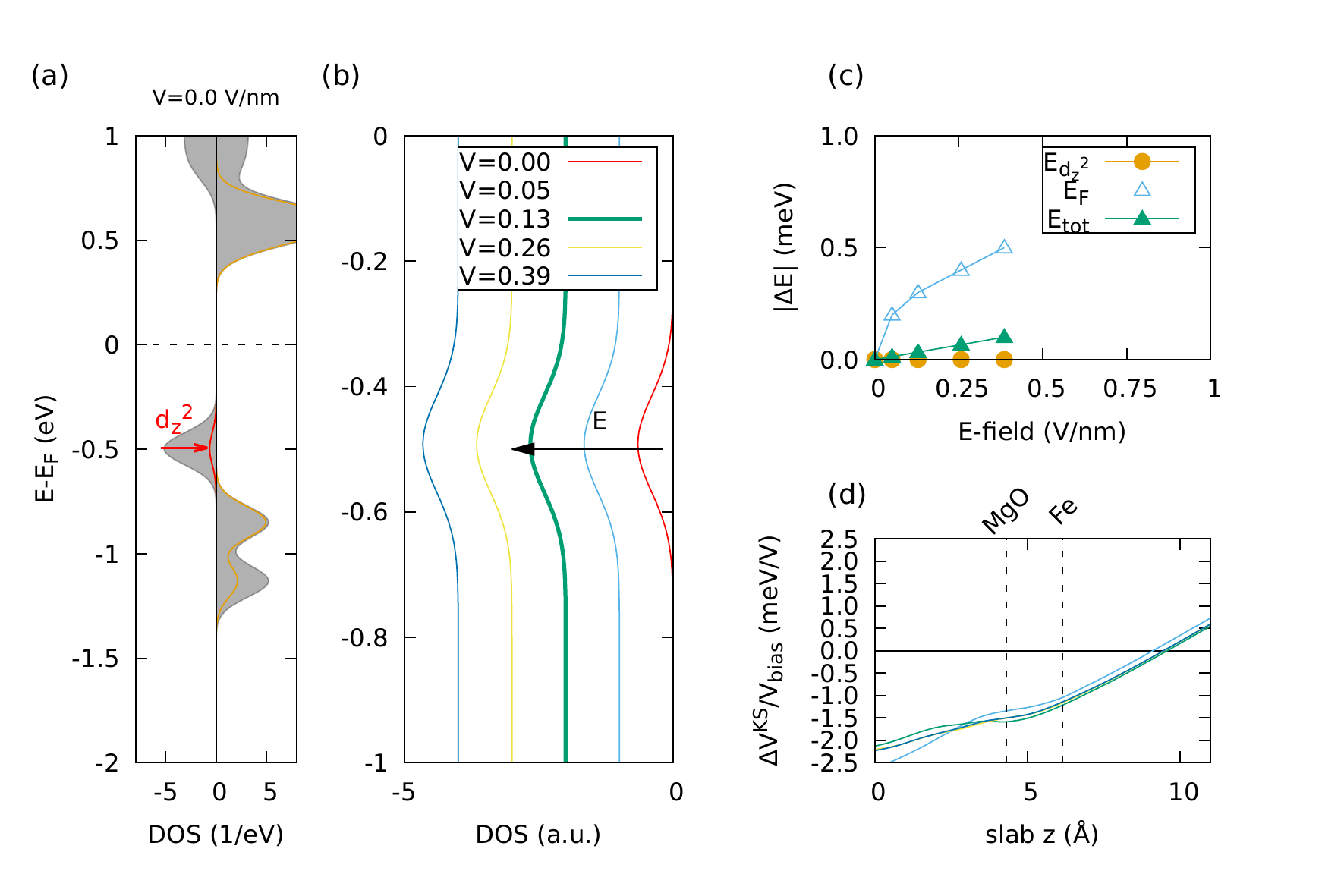}
\caption{Effect of an applied electric field on the band structure of Fe on MgO. (a) shows the LDOS of the Fe:3d manifold superimposed over the total DOS (grey). The level closest to the Fermi level (at -0.5 eV) has mostly $d_z^2$ character. The shift of this level with applied homogeneous electric field is shown in (b). For the fields relevant in the experiment (roughly 0.1 V/nm), this level does not exhibit any significant shift. In (c) the shift of the $d_z^2$, the Fermi level $E_F$, and the total energy $E_{tot}$ are shown. In all cases the shift up to fields of 0.4 V/nm are negligible and will not lead to a change of our multiplet results. (d) shows the in-plane averaged Kohn-Sham potential relative to the potential of the case with zero field divided by the local electric field amplitude. The position of the substrate and adatom are marked by vertical dashed lines.}
\label{fig:sup:pdos_field}
\end{figure*}

When calculating the force of the external electric field on the adatom (the case of Fe on MgO is shown in Figure~\ref{fig:sup:force_Fe}), it is found that an external field up to 5 V/nm (corresponding to 0.5 V applied over a distance of 1~\AA~ and assuming a plate capacitor model) does not exert a significant force on the adatom. Further, the external field does not significantly shift the Kohn-Sham levels of the adatom close to the Fermi level, 
as shown in Figure~\ref{fig:sup:pdos_field}. All findings indicate that electrical field strengths in the STM junction do not significantly perturb the electronic state of the adatom as previously discussed in Ref.\cite{ReinaGalvez2019}

\end{document}